\documentclass[prl, twocolumn, showpacs, superscriptaddress]{revtex4-1}
\usepackage{amsmath, amssymb}
\usepackage{graphicx}
\usepackage{setspace}
\usepackage{color}
\renewcommand{\section}[1]{{\par\it #1.---}\ignorespaces}

\begin{document}
\title{Non-monotonic field dependence of Kondo conductance in a single-electron transistor driven by microwave field}
\author{Zhan Cao}
\affiliation{Center for Interdisciplinary Studies $\&$ Key Laboratory for
Magnetism and Magnetic Materials of the MoE, Lanzhou University, Lanzhou 730000, China}
\author{Chen Cheng}
\affiliation{Center for Interdisciplinary Studies $\&$ Key Laboratory for
Magnetism and Magnetic Materials of the MoE, Lanzhou University, Lanzhou 730000, China}
\author{Fu-Zhou Chen}
\affiliation{Center for Interdisciplinary Studies $\&$ Key Laboratory for
Magnetism and Magnetic Materials of the MoE, Lanzhou University, Lanzhou 730000, China}
\author{Hong-Gang Luo}
\affiliation{Center for Interdisciplinary Studies $\&$ Key Laboratory for
Magnetism and Magnetic Materials of the MoE, Lanzhou University, Lanzhou 730000, China}
\affiliation{Beijing Computational Science Research Center, Beijing 100084, China}

\pacs{73.23.-b, 73.23.Hk, 72.15.Qm, 78.70.Gq}

\begin{abstract}
The interplay between magnetic field and microwave applied in a single-electron transistor(SET) has a profound influence on the Kondo effect, as shown in a recent experiment[B. Hemingway, S. Herbert, M. Melloch and A. Kogan, arXiv:1304.0037(2013)]. For a given microwave frequency, the Kondo differential conductance shows a non-monotonic magnetic field dependence, and a very sharp peak is observed for certain field applied. Additionally, the microwave frequency is found to be larger of about one order than the corresponding Zeeman energy. These two features are not understood in the current theory. Here we propose a phenomenological mechanism to explain these observations. When both magnetic field and microwave are applied in the SET, if the frequency matches the (renormalized) Zeeman energy, it is assumed that the microwave is able to induce spin-flip in the SET, which leads to two consequences. One is the dot level shifts down and the other is the renormalization of the Zeeman energy. This picture can not only explain qualitatively the main findings in the experiment but also further stimulate the related experimental study of the dynamic response of the Kondo effect in out-of-equilibrium devices.
\end{abstract}
\maketitle

\section{Introduction} \label{intr}
The Kondo physics is one of fundamental issues in exploring many-body correlations\cite{Kondo1964, Hewson1993}, which originates from the screening of local moment by conduction electrons in metal. Theoretical prediction \cite{Ng1988, Glazman1988} and its physical realization \cite{Goldhaber-Gordon1998, Cronenwett1998, Schmid1998} of the Kondo effect in a single-electron transistor(SET) provide a great chance to investigate intriguing features of the Kondo physics by tuning various controllable parameters such as gate voltage, source-drain voltage, the magnetic field, and so on. The existence of the Kondo resonance near the Fermi level in the SET affects dramatically the transport of the device, and a unitary conductance in the Coulomb-blockade regime can be reached\cite{Cronenwett1998}. In addition, this set-up is also an ideal platform to study non-equilibrium many-body correlation if these parameters are changed in time, as explored extensively in experiments \cite{Elzerman2000, Kogan2004, Delbecq2011} and theories \cite{Meir1993, Ng1993, Hettler1994,Schiller1996,Goldin1998,Lopez1998,Kaminski1999,Nordlander2000,Nguyen2012}. When the SET is irradiated with microwave, photon-induced Kondo satellites have been observed experimentally when appropriate microwave frequency and amplitude were applied\cite{Kogan2004}.

Very recently, Hemingway \textit{et al.} \cite{Hemingway2013} measured the time-averaged differential conductance of the SET and observed some novel transport behaviors when magnetic field and microwave are applied at the same time. They found that i) at zero magnetic field the microwave applied suppresses the Kondo effect when the photon energy is comparable or greater than the Kondo temperature, which is in agreement with available theoretical prediction \cite{Kaminski1999}; ii) at finite magnetic field the Kondo conductance changes non-monotonically with magnetic field for a given microwave frequency and the conductance shows a very sharp peak as a function of magnetic field, this non-monotonic behavior is absent at low frequencies; and iii) the microwave frequency is larger of about one order than the corresponding Zeeman energy applied while the anomalous non-monotonic behavior is observed. Since the first observation is well understood by available theory \cite{Kaminski1999} and the latter two features are novel but can currently not be understood, below we focus on the case of finite magnetic field and clarify the latter two features.

\begin{figure}[tbp]
\includegraphics[width=0.9\columnwidth]{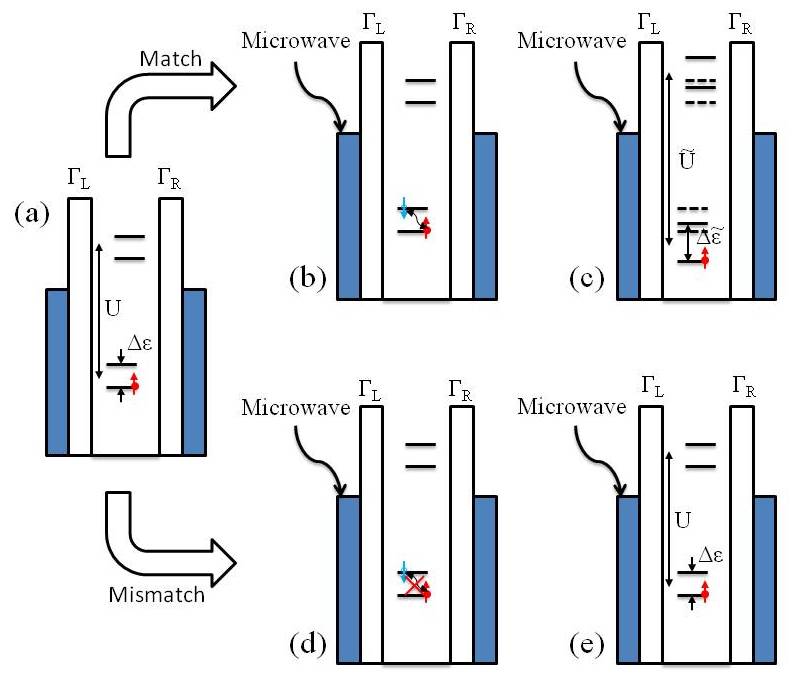}
\caption {Schematic dot levels in the SET when both magnetic field and microwave are applied. (a) The dot level Zeeman splitting without microwave irradiation. $\Delta \varepsilon$ and $U$ are the Zeeman energy and the Coulomb repulsion, respectively. (b) $\&$ (c) The photon energy matches the (renormalized) Zeeman energy. A spin-flip transition in the SET is induced and two renormalized effects occur, namely, the downshift of the dot level and the increase of the effective Zeeman energy (due to the experimental observation here). The dashed lines are unrenormalized levels. $\Delta\tilde\varepsilon$ and $\tilde U$ are renormalized Zeeman energy and Coulomb repulsion, respectively. (d) $\&$ (e) The photon energy does not match the Zeeman energy, thus the spin-flip transition is prohibited. The dot levels remain unchanged. }\label{fig1}
\end{figure}

First of all, we present our basic picture. Fig. \ref{fig1} shows a schematic energy level diagram of the SET. In the absence of microwave [Fig.\ref{fig1}(a)], the dot level splits due to the Zeeman energy $\Delta \varepsilon = g\mu_B B$ where $g$ is Land\'e factor, $\mu_B$ Bohr magneton and $B$ magnetic field applied. When the SET is irradiated with microwave, some parameters, for example, the bias voltage and dot-leads tunneling in the SET, may vary in time, which lead to many interesting features such as the satellites\cite{Kaminski1999,Kogan2004,Nguyen2012}. However, in the experiment \cite{Hemingway2013} these satellites have not been observed and therefore here we will focus on another possible effect, namely, the spin-flip transition induced by microwave field, which has been extensively studied in double quantum dots system\cite{Oosterkamp1998,van der Wiel2002, Petta2004,Nowack2007,Laird2007,Pioro-Ladriere2008,Petersson2009,Schreiber2011} but has not been explored in the present set-up. Here whether the photon energy matches the Zeeman energy or not leads to quite different physics. If match, the microwave can induce a spin-flip transition, as shown in Fig. \ref{fig1}(b), as a consequence of this transition the dot levels are renormalized[solid lines in Fig. \ref{fig1}(c)]. Two effects are observed. The dot levels for both spin shift down and at the same time, the Zeeman splitting increases effectively(due to the experimental observation, as discussed later). If not match, the spin-flip transition can not happen. As a result, the dot levels remain almost unchanged, as shown in Fig. \ref{fig1}(d,e). As shown below, for a given microwave frequency in experiment, the existence of a sharp peak of the Kondo conductance as a function of magnetic field applied can be simply attributed to the resonance between the frequency and magnetic field. This is because that in this case the dot energy is renormalized to more deeper level and thus the Kondo effect is strengthened effectively. In this picture we clarify qualitatively the anomalous features observed experimentally. The result can further stimulate the related experimental study of the dynamic response of the Kondo effect in non-equilibrium devices.

\section{Model and renormalized dot levels}
In the absence of microwave irradiation, the SET can be described by the following Hamiltonian
\begin{eqnarray}
&& H_{\text{SET}} = \sum_{k\sigma,\alpha = L,R} \varepsilon_{k\alpha\sigma}c_{k\alpha\sigma}^{\dag}c_{k\alpha\sigma} + \sum_{\sigma}\varepsilon_{d\sigma} d^\dagger_\sigma d_\sigma + U n_{d\uparrow}n_{d\downarrow} \nonumber\\
&&\hspace{2cm} + \sum_{k\alpha\sigma}\left( V_{\alpha} d_{\sigma}^{\dagger}c_{k\alpha\sigma} + h.c.\right), \label{eq1}
\end{eqnarray}
where the first term represents the left(L) and right(R) leads, the successive two terms denote dot Hamiltonian and the last one is the hybridization between the dot and leads. $\varepsilon_{d\sigma} = \varepsilon_d + \frac{\sigma}{2}g\mu_B B$ and $U$ are the dot levels with spin $\sigma = \pm(\uparrow,\downarrow)$ and the on-site Coulomb repulsion.

In the presence of microwave irradiation with frequency $f$, in principle the device is in out-of-equilibrium and many parameters may vary in time. As mentioned above, here we only focus on the spin-flip transition induced by the microwave field, which can be captured by the following effective Hamiltonian
\begin{equation}
H_{\text{p-d}} = \omega_{p}a^{\dagger}a + \lambda(a^{\dagger} + a)\sum_{\sigma}d_{\sigma}^{\dagger}d_{\bar{\sigma}},\label{eq2}
\end{equation}
where $\omega_p = hf$($h$: Planck constant) and $\lambda$ is the coupling strength between the dot and the photon, which depends on the match between the photon energy and the Zeeman energy, as discussed below. Thus the total Hamiltonian is given by
\begin{equation}
H = H_{\text{SET}} + H_{\text{p-d}}. \label{eq3}
\end{equation}
To derive the renormalized dot level, we first apply the canonical transformation $H\rightarrow \tilde H = e^S H e^{-S}$
with $S = \frac{\lambda}{\omega_p}(a^\dagger - a)\sum_\sigma d^\dagger_\sigma d_{\bar\sigma}$, one can obtain the following effective Hamiltonian (see Sec. I in Supplemental Materials\cite{note})
\begin{eqnarray}
&& \tilde H \approx \sum_{k\sigma,\alpha = L,R} \varepsilon_{k\alpha\sigma}c_{k\alpha\sigma}^{\dag}c_{k\alpha\sigma} + \sum_{\sigma}\tilde{\varepsilon}_{d\sigma} d^\dagger_\sigma d_\sigma + \tilde U n_{d\uparrow}n_{d\downarrow} \nonumber\\
&&\hspace{2cm} + \sum_{k\alpha\sigma}\left( \tilde V_{\alpha} d_{\sigma}^{\dagger}c_{k\alpha\sigma} + h.c.\right), \label{eq1-a}
\end{eqnarray}
where
\begin{equation}
\tilde\varepsilon_{d\sigma} = \varepsilon_d - \frac{\lambda^2}{\omega_p} +\frac{\sigma}{2}\Omega_{\lambda}\Delta\varepsilon,\label{eq4}
\end{equation}
and $\tilde U = U + \frac{2\lambda^2}{\omega_p}$, $\tilde V_{\alpha} = \Omega'_{\lambda} V_\alpha$, $\Delta\varepsilon=g\mu_B B$. Here $\Omega_{\lambda}$ and $\Omega'_{\lambda}$ are renormalized parameters given in Sec. I in Supplemental Materials \cite{note}.

Equations (\ref{eq1-a}) and (\ref{eq4}) are our central results of the present work. In comparison to Eq. (\ref{eq3}), the influence of the photons is renormalized into the quantum dot parameters, as shown in $\tilde \varepsilon_{d\sigma}, \tilde U$, and $\tilde V_\alpha$. When $\lambda = 0$, it is easy to show that all renormalized parameters go back to the original ones and Eq. (\ref{eq1}) is recovered. From Eq. (\ref{eq4}) the coupling with photons produces two important consequences. One is that the dot levels for both spin have a downshift with $\lambda^2/\omega_p$, whose consequence is to strengthen Kondo effect. The other is the renormalization of the Zeeman energy which can not be evaluated exactly here but can be fixed phenomenologically by the experiment (see below).

It should be emphasized that the above renormalization happens only under the resonance condition, namely, the microwave frequency match the renormalized Zeeman energy, namely, $\lambda = \lambda_0\delta(\omega_p - \Omega_\lambda\Delta\varepsilon)$, where $\lambda_0$ is the coupling strength dependent of the microwave frequency applied. In calculation, this can be expressed by a Lorentzian function as follow in a dimensionless form
\begin{equation}
\lambda = \frac{\lambda_0}{\pi}\frac{\beta}{(\omega_p - \Omega_\lambda\Delta\varepsilon)^2 + \beta^2}, \label{eq5}
\end{equation}
where $\beta\rightarrow 0^{+}$ is the width of the resonance. As argued in Sec. I of Supplemental Materials, here we take phenomenologically $\Omega_\lambda\sim 7$, which is determined by experimental observation.
\begin{figure}[tbp]
\begin{center}
\includegraphics[width=0.9\columnwidth]{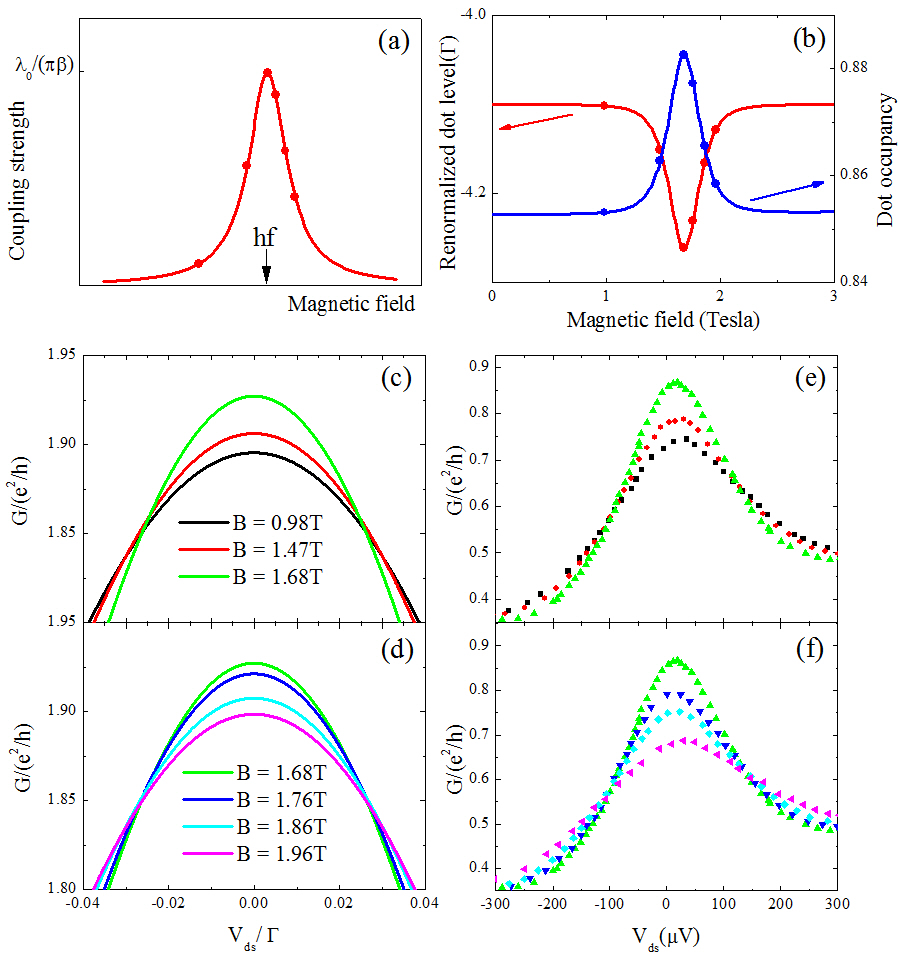}
\caption{(Color online) (a) The coupling strength $\lambda$ as a function of magnetic field applied for a given frequency $f = 34.1$GHz. The dots represent the coupling strengths corresponding to magnetic fields taken in experiment[same in Fig. \ref{fig2}(b)]. (b) The renormalized spin-independent dot level $\tilde\varepsilon_d$ and the dot occupancy as functions of magnetic fields. (c) $\&$ (d) Theoretical Kondo conductance for different magnetic fields at zero temperature. (e) $\&$ (f) Experimental Kondo conductance for different magnetic fields (reproduced from Ref. \cite{Hemingway2013}). The parameters used are $\varepsilon_d = -4.1\Gamma$, $\lambda_0 = 3\times 10^{-4} \Gamma$, $\beta=2\times 10^{-3} \Gamma$, the half-bandwidth $D = 100\Gamma$ and the renormalized broadening $\Gamma = 10$meV. The Land\'e factor $g = 0.207$ and $\mu_B = 58\mu eV/T$.}\label{fig2}
\end{center}
\end{figure}
\begin{figure}[tbp]
\begin{center}
\includegraphics[width=0.9\columnwidth]{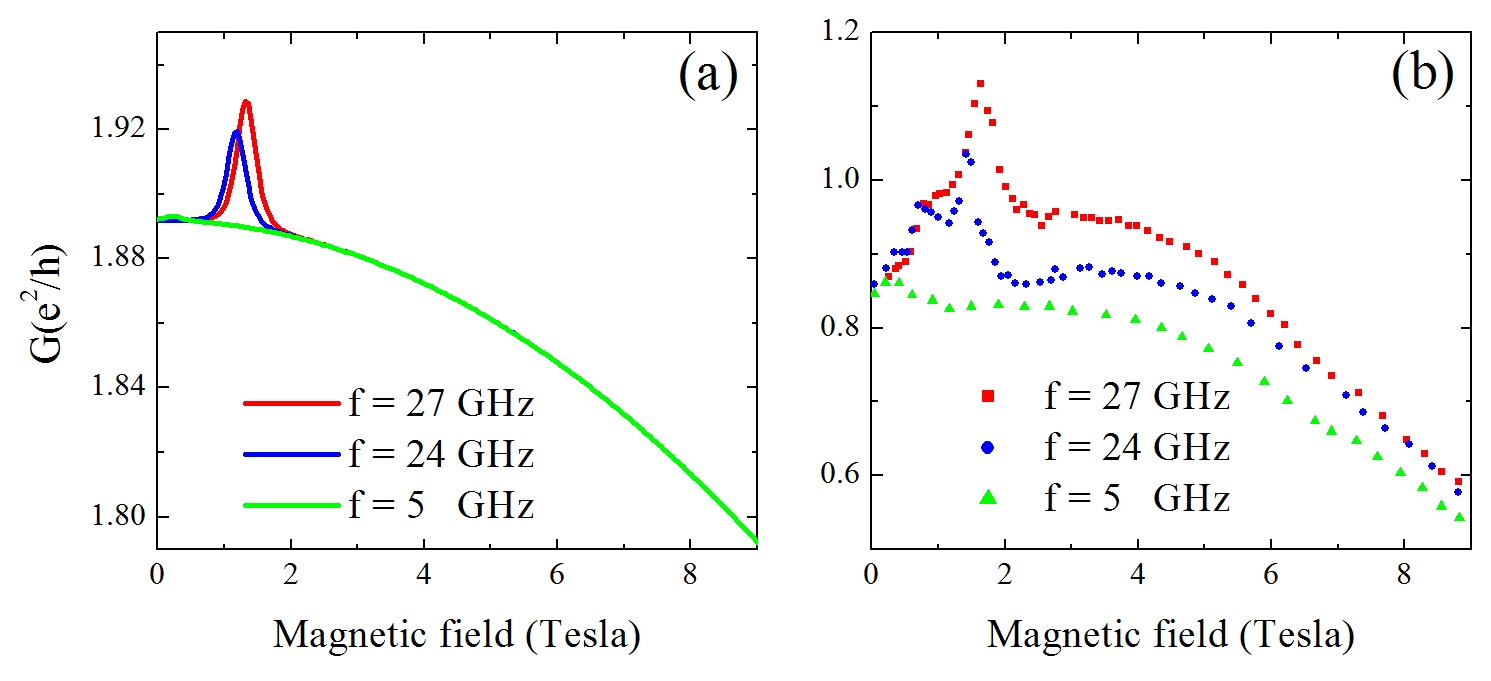}
\caption{(Color online) (a) Theoretical and (b) experimental differential conductance at the Fermi level as function of magnetic field applied for three different microwave frequency. The parameters used are $\lambda_0 = (2.8, 2.2, 0.2) \times 10^{-4} \Gamma$ for $f=(27, 24, 5)$ GHz in order to fit the heights of the conductance for different frequency. The other parameters used are the same as those in Fig. \ref{fig2}.}\label{fig3}
\end{center}
\end{figure}

\section{Comparison with experiment} Below we address the experiment on the following two aspects. One is to starting from the effective Hamiltonian Eq. (\ref{eq4}), in which the effect of the microwave irradiation is already included as renormalized parameters. The other is to fit directly the experimental data by the Kondo resonance and to extract the Kondo temperature as a function of magnetic fields applied.

To calculate the differential conductance, we use the Keldysh formalism\cite{Haug2008} and the slave-boson mean field method (see Sec. II and III in Supplemental Materials\cite{note}, respectively). Near the Fermi level, the slave-boson mean-field is sufficient to capture the essential physics of the Kondo effect. From these calculations experimental observations can be reproduced qualitatively, as discussed below. For a given microwave frequency $f = 34.1 $GHz(obtained by $hf = \Omega_\lambda g\mu_B B$, $\Omega_\lambda \sim 7$, $g=0.207$, and $B = 1.68$T\cite{Hemingway2013}), when magnetic field is varied, the coupling strength changes according to if match between the frequency and magnetic field applied or not, as shown in Fig. \ref{fig2}(a). For different magnetic fields [denoted by dots in Fig.\ref{fig2}(a)] the differential conductance is presented in Fig. \ref{fig2}(c) and (d) at zero temperature. With increasing magnetic field one notes that the height of the conductance peak increases and reaches a maximal value at $B = 1.68$T, where the (renormalized) Zeeman energy matches the microwave frequency and the dot occupancy also reaches maximum due to the downshift of the dot level, as shown in Fig. \ref{fig2}(b). Further increasing magnetic field, the height of the peak decreases. Thus the height of the Kondo conductance is non-monotonic as a function of magnetic field, which is qualitatively consistent with the experimental observations, as shown in Fig. \ref{fig2}(e,f). One also notes that the width of the peaks in Fig. \ref{fig2}(c) and (d) is much broader than that in experiment, which is due to the approximate slave-boson mean field method. In the above discussion the Zeeman splitting is invisible due to small magnetic field.

To further confirm this observation, in Fig. \ref{fig3}(a) we present the differential conductance at the Fermi level for different microwave frequency and compare directly to the experimental results. The agreement with the experiment is qualitatively good. Thus one can conclude that the essential physics of the non-monotonic field dependence of the Kondo conductance is that the dot level has a downshift due to the renormalization effect if the photon energy matches the renormalized Zeeman energy, as mentioned above. In addition, the experiment also indicates that for high frequency, besides a very sharp peak, at low magnetic fields, about half and one quarter there are obvious peaks and/or shoulder structures, which is further explained qualitatively in Sec. IV of Supplemental Materials\cite{note}.

Though the above discussion captures the essential physics of the experimental observation in the SET at finite magnetic field, there still exists quantitative difference between the calculation and the experiment due to some unknown parameters such as the dot level, tunneling matrix element and the approximate method used. In the following we focus on the experimental data. Since the SET is tuned to be in Kondo regime, the feature near Fermi level is nothing but the Kondo resonance, which can be simplified as
\begin{equation}
T_d(\omega) \approx \frac{1}{\omega - \varepsilon_K + i T_K},\label{eq6}
\end{equation}
where $\varepsilon_K$ is the location of the Kondo resonance with half-width $T_K$, namely, the Kondo temperature. Here we still neglect the Zeeman splitting due to small magnetic field. This is justified by the experiment, in which no sizable Zeeman splitting is observed except for $B = 8.8$T \cite{Hemingway2013}. Phenomenologically, we use the following Fano formula to fit the experimental data.
\begin{equation}
G = G_0 - \rho_0[(q^2 - 1) \text{Im}T_d(\omega) - 2 q \text{Re} T_d(\omega)],
\end{equation}
 where $q$ is Fano asymmetric factor \cite{Fano1961, Luo2004} to describe the asymmetry of the differential conductance and $G_0$, $\rho_0$ are background parameters. The result is presented in Fig. \ref{fig4}(a), which shows well fitting (the fitting parameters are presented in Sec. V in Supplemental Materials\cite{note}). Fig. \ref{fig4}(b) shows the Kondo temperature extracted from the fitting for different magnetic fields. The logarithm scale presents the field dependence of the dot level up to a constant. This can be obtained by the exact Kondo temperature expression $T_K \propto \exp[\pi(\varepsilon_d - \varepsilon_F)/2\Delta]$\cite{Haldane1978}, where $\Delta$ is the tunneling matrix element and $\varepsilon_F$ is the Fermi level. For convenience to compare, in Fig. \ref{fig4}(c) we again present the renormalized dot level as a function of magnetic field for three different microwave frequency. The comparison further supports our theoretical picture.
\begin{figure}[tbp]
\includegraphics[width=0.9\columnwidth]{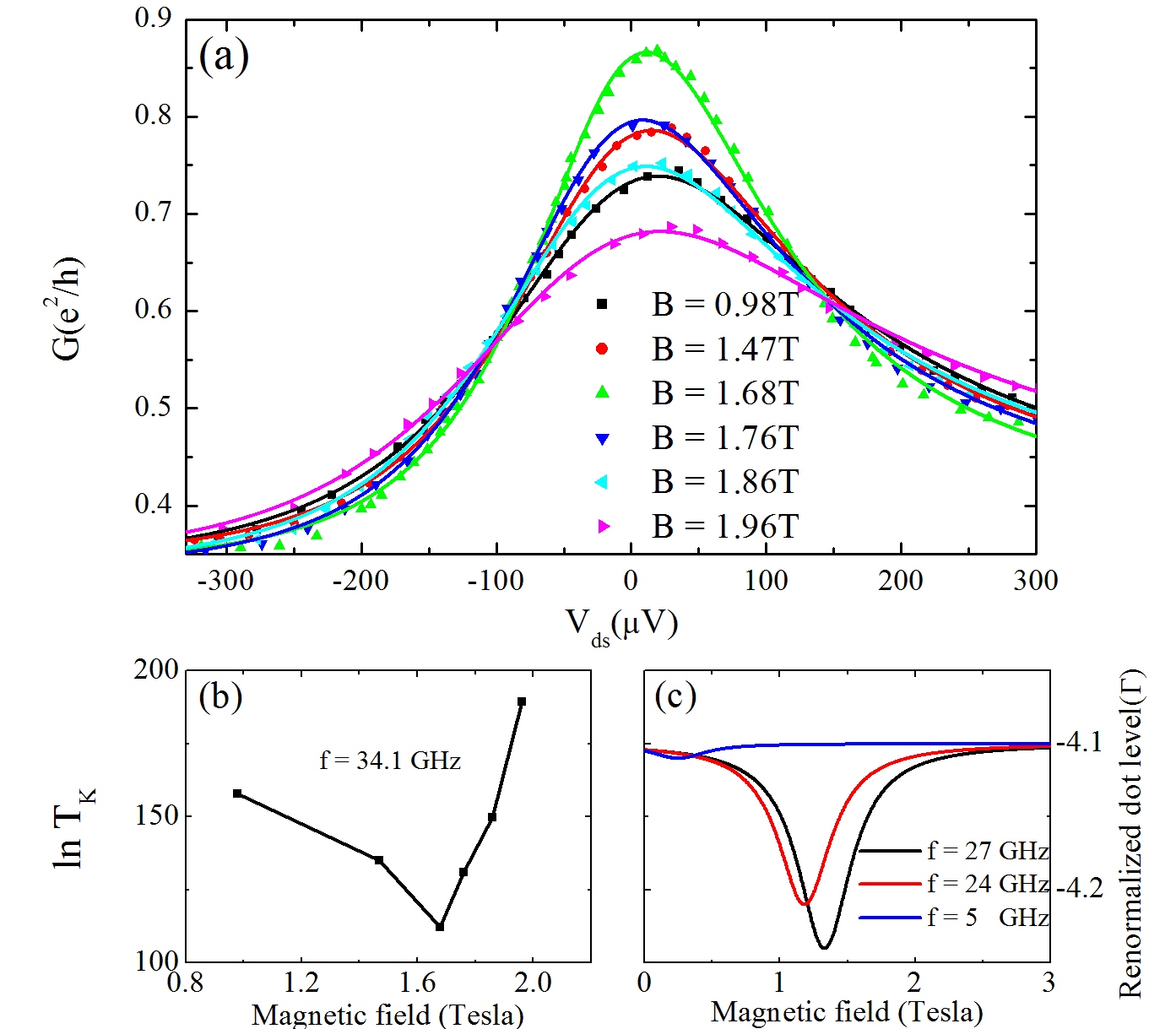}
\caption {(Color online) (a) Experimental (scatters) differential conductance as function of magnetic field and their theoretical fitting (solid lines). (b) The logarithm of the Kondo temperature extracted from the fitting, which is proportional to the dot level. (c) The renormalized dot levels as function of magnetic field for three different microwave frequency.}\label{fig4}
\end{figure}

\section{Discussion and outlook}
In Introduction, we mentioned that the non-equilibrium Kondo physics has been intensively investigated in the previous theoretical works \cite{Meir1993, Ng1993, Hettler1994,Schiller1996,Goldin1998,Lopez1998,Kaminski1999,Nordlander2000,Nguyen2012}, most of which focused on the time-varying bias voltages. In this case a striking phenomenon predicted is the satellite peaks in the dependence of the differential conductance on the dc bias voltage, which has been confirmed experimentally in 2004 by Kastner's group \cite{Kogan2004}. In particular, very recently, Nguyen \cite{Nguyen2012} has studied in detail dynamic response of an SET in a magnetic field irradiated with microwaves. They explored two fold effect of the microwave, one is the oscillation in voltage with microwave frequency $\Omega$ and the other is the oscillation in the coupling parameter with frequency $\Omega/p(p\in N)$. While the former has been well studied in the previous works, the latter one leads to the central result of that work, namely, the satellite peak splitting \cite{Nguyen2012}. It is a delicate situation to observe the satellite peaks, which requires properly the frequency and amplitude of the microwaves\cite{Kogan2004}. However, in the recent experiment carried out by Kogan's group \cite{Hemingway2013}, neither the satellite peaks nor the satellite peak splitting have been observed. Instead, the differential conductance shows a non-monotonic field dependence for a given microwave frequency greater than the Kondo temperature. This result indicates that the physics behind the experiment does not fall into the framework studied in the previous works \cite{Nguyen2012}. Several spin-flip transition mechanisms in quantum dot has been explored by Khaetskii and Nazarov \cite{Khaetskii2001} but they are irrelevant to microwave excitation mechanism discussed here.

By renormalization effect on the Zeeman splitting one can understand why the microwave frequency disagrees quantitatively with the previous theoretical prediction\cite{Schiller1996} $\Delta\varepsilon = hf$. However, we are unable to calculate exactly the renormalization factor due to the coupling between the electrons and the photons. Therefore, we take this renormalization factor from the experiment. A further problem left is not answered is that the renormalization factor is almost a constant of about $7$ observed in the experiment for both SETs\cite{Hemingway2013}. Irrespective of the detail, the factor should be related to the the photon population which can be crudely estimated at temperature $T$ by $n_p = 1/\left(\text{e}^{hf/k_BT}-1\right)$. It is noted that in experiment the base electron temperature in both SETs is about $T = 70$mK, which is kept fixed. It is instructive to change the temperature in the cavity thus one can check if the renormalization factor changes or not. This should be tested by experiment. On the other hand, it is also interesting to explore possible influence of the interplay between the microwave excitation and the possible spin-flip transition mechanisms \cite{Khaetskii2001} on the non-equilibrium transport of the SET.

In summary, we have proposed a phenomenological mechanism to explain the non-monotonic magnetic field dependence of the Kondo conductance observed in a recent experiment, which can not be understood in the current theory. The essential physics is the renormalization effects attributed to the spin-flip transition induced by the microwave when the photon energy matches the renormalized Zeeman energy. The result sheds a light on the transport behavior of the related devices.

\section{Acknowledgments}
The authors thank S. Y. Cho and T.-F. Fang for valuable discussion. The work is partly supported by the programs for NSFC, PCSIRT (Grant No. IRT1251), the national program for basic research and the Fundamental Research Funds for the Central Universities of China.

\end{document}


\title{Supplemental materials}
\author{Zhan Cao, Chen Cheng, Fu-Zhou Chen, and Hong-Gang Luo}
\maketitle

\section{DERIVATION OF THE EFFECTIVE HAMILTONIAN}
Similar to the scheme extensively adopted in studying the Anderson-Holstein (AH) model\cite{David2002, Mahan2000}, we apply a generalized Lang-Firsov transformation to each operator $O$ presented in the original Hamiltonian (1-3) in the main text as
\begin{equation}
e^{S}Oe^{-S}=O+\left[  S,O\right]  +\frac{1}{2!}\left[  S,\left[  S,O\right]
\right]  +\frac{1}{3!}\left[  S,\left[  S,\left[  S,O\right]  \right]
\right]  +... ,
\end{equation}
where $S=\frac{\lambda}{\omega_{p}}\left(  a^{+}-a\right)  \sum_{\sigma}d_{\sigma}^{\dagger}d_{\bar{\sigma}}$. It is no difficult to obtain
\begin{eqnarray}
&& e^{S}d_{\sigma}e^{-S}  = X d_{\sigma}-Y d_{\bar{\sigma}},\\
&& e^{S}d_{\sigma}^{\dagger}e^{-S} = X d_{\sigma}^{\dagger}+Y d_{\bar{\sigma}}^{\dagger}.
\end{eqnarray}
Here $X, Y$ are given by
\begin{eqnarray}
&&X =\frac{1}{2}\left[F\left(\frac{\lambda}{\omega_{p}}\right)+F\left(-\frac{\lambda}{\omega_{p}}\right)\right], \\
&&Y=\frac{1}{2}\left[F\left(\frac{\lambda}{\omega_{p}}\right)-F\left(-\frac{\lambda}{\omega_{p}}\right)\right],
\end{eqnarray}
where $F(\frac{\lambda}{\omega_{p}})=e^{\frac{\lambda}{\omega_{p}}\left(  a^{\dagger}-a\right)  }$. It's easy to show that $X$ and $Y$ are Hermitian and anti-Hermitian operators, respectively, which satisfy $X^{\dag}=X$ and $Y^{\dag}=-Y$. Thus we have
\begin{eqnarray}
&&X^{2}=\frac{1}{4}\left[  F\left(\frac{2\lambda}{\omega_{p}}\right)+F\left(-\frac{2\lambda}{\omega_{p}}\right)\right]+\frac{1}{2},\\
&&Y^{2}=\frac{1}{4}\left[  F\left(\frac{2\lambda}{\omega_{p}}\right)+F\left(-\frac{2\lambda}{\omega_{p}}\right)\right]-\frac{1}{2},\\
&&XY=\frac{1}{4}\left[     F\left(\frac{2\lambda}{\omega_{p}}\right)-F\left(-\frac{2\lambda}{\omega_{p}}\right)\right],\\
&&X^{2}-Y^{2}=1.
\end{eqnarray}
For the creation and annihilation operators of photon, we have
\begin{eqnarray}
e^{S}ae^{-S} &  =a-\frac{\lambda}{\omega_{p}}\sum_{\sigma}d_{\sigma}^{\dagger}d_{\bar{\sigma}},\\
e^{S}a^{\dagger}e^{-S} &  =a^{\dagger}-\frac{\lambda}{\omega_{p}}\sum_{\sigma}d_{\sigma}^{\dagger}d_{\bar{\sigma}}.
\end{eqnarray}
Since the relation $e^{-S}e^{S}=1$ that $e^{S}MNe^{-S}=e^{S}Me^{-S}e^{S}Ne^{-S}$. We can readily obtain the transformed forms of each ingredient of the original Hamiltonian as follows
\begin{eqnarray}
\hspace{-0.2cm}e^{S}\sum_{\sigma}\varepsilon_{d\sigma}n_{d\sigma}e^{-S} &  =\sum_{\sigma}\left[  {\varepsilon}_{d}+\frac{\sigma}{2}\left(  X^{2}+Y^{2}\right)  \Delta\varepsilon\right]  n_{d\sigma} \notag\\
&-XY\Delta\varepsilon\sum_{\sigma}\sigma d_{\sigma}^{\dagger}d_{\bar{\sigma}},
\end{eqnarray}
\begin{eqnarray}
e^{S}Un_{d\uparrow}n_{d\downarrow}e^{-S}=Un_{d\uparrow}n_{d\downarrow},\\
e^{S}\sum_{\sigma}d_{\sigma}^{\dagger}d_{\bar{\sigma}}e^{-S}=\sum_{\sigma}d_{\sigma}^{\dagger}d_{\bar{\sigma}},
\end{eqnarray}
\begin{eqnarray}
& e^{S}\lambda\left(  a^{+}+a\right)  \sum_{\sigma}d_{\sigma}^{\dagger}d_{\bar{\sigma}}e^{-S}= \lambda\left(  a^{\dagger}+a\right)  \sum_{\sigma}d_{\sigma}^{\dagger}d_{\bar{\sigma}} \notag\\
& -\frac{2\lambda^2}{\omega_p}\sum_{\sigma}n_{d\sigma}+\frac{2\lambda^2}{\omega_p}\sum_{\sigma}n_{d\sigma}n_{d\bar{\sigma}},
\end{eqnarray}

\begin{eqnarray}
&e^{S}\omega_{p}a^{\dagger}ae^{-S}=\omega_{p}a^{\dagger}a-\lambda\left(  a^{\dagger}+a\right)  \sum_{\sigma}d_{\sigma}^{\dagger}d_{\bar{\sigma}} \notag\\
&+\frac{\lambda^2}{\omega_p}\sum_{\sigma}n_{d\sigma}-\frac{\lambda^2}{\omega_p}\sum_{\sigma}n_{d\sigma}n_{d\bar{\sigma}}.
\end{eqnarray}
Summing over above terms leads to a completely equivalent Hamiltonian
\begin{eqnarray}
&&\tilde{H}=\sum_{k\sigma,\alpha}\varepsilon_{k\sigma\alpha}c_{k\sigma\alpha}^{\dagger}c_{k\sigma\alpha}+\sum_{\sigma}\tilde{\varepsilon}_{d\sigma}n_{d\sigma}
+\tilde U n_{d\uparrow}n_{d\downarrow} +\omega_{p}a^{\dagger}a \notag\\
&&-XY\Delta\varepsilon\sum_{\sigma}\sigma d_{\sigma}^{\dagger}d_{\bar{\sigma}} +\sum_{k\sigma\alpha}\left[ V_{\alpha}\left(X d_{\sigma}^{\dagger}+Y d_{\bar{\sigma}}^{\dagger}\right)c_{k\alpha\sigma}+h.c.\right], \label{16}
\end{eqnarray}
where $\tilde{\varepsilon}_{d\sigma} = {\varepsilon}_{d}-\frac{\lambda^2}{\omega_p}+\frac{\sigma}{2}\left(  X^{2}+Y^{2}\right)\Delta\varepsilon$ and $\tilde U = U+\frac{2\lambda^2}{\omega_p} $.
This Hamiltonian implies a series of renormalization effect on the model parameters such as the dot level, Zeeman energy, Coulomb repulsion and coupling strength between dot and conducting leads.

Formally, the photon Hamiltonian is decoupled from the electron Hamiltonian. However, this is not true due to the presence of operators $X$ and $Y$, which contain the photon operator. In the mean-field level, one can treat $\Omega_\lambda = \langle X^2 + Y^2\rangle$ as renormalization factor of the Zeeman energy, where $\langle \cdot \rangle = \langle \cdot \rangle_{\tilde H}$. However, due to the coupling between the photons and the electrons, it is difficult to evaluate exactly the expectation for the Hamiltonian $\tilde H$. Here we do not pursue the analytical evaluation of this renormalization factor but follow the experiment observation to take phenomenologically this renormalization factor as $\Omega_\lambda \sim 7$ in our calculation. In this case, the following Hamiltonian is taken as our starting point
\begin{eqnarray}
&& \tilde{H}=\sum_{k\sigma,\alpha}\varepsilon_{k\sigma\alpha}c_{k\sigma\alpha}^{\dagger}c_{k\sigma\alpha}+\sum_{\sigma}\tilde{\varepsilon}_{d\sigma}n_{d\sigma}
+\tilde U n_{d\uparrow}n_{d\downarrow} \nonumber\\
&& \hspace{2cm} +\sum_{k\sigma\alpha}\left[\tilde{V}_{\alpha} d_{\sigma}^{\dagger}c_{k\sigma\alpha}+h.c.\right], \label{16b}
\end{eqnarray}
where $\tilde{\varepsilon}_{d\sigma} = \varepsilon_d - \frac{\lambda^2}{\omega_p} +\frac{\sigma}{2}\Omega_\lambda \Delta \varepsilon$, $\tilde U = U + \frac{2\lambda^2}{\omega_p}$, and $\tilde{V}_{\alpha} = \Omega'_\lambda V_\alpha$ with $\Omega'_\lambda = \langle X \rangle_{\tilde H}$. Here we neglect $\langle Y\rangle_{\tilde H}$ and $\langle X Y\rangle_{\tilde H}$, which are believed to be vanishing small due to $Y$'s definition. Actually, one can show $\langle F(\frac{\lambda}{\omega_p})\rangle_{H_{p}}=\langle F(-\frac{\lambda}{\omega_p})\rangle_{H_{p}}$, here $H_{p} = \omega_{p}a^{\dagger}a$ is the Hamiltonian of photons.

\section{BRIEF INTRODUCTION TO KELDYSH FORMALISM}
In order to study the transport property of the model Eq. (\ref{16b}), we use the usual Keldysh formalism\cite{Haug2008}. According to Ref. \cite{Meir1992}, the electronic current formula is given by
\begin{equation}
I=\frac{2e}{\pi\hbar}\frac{\Gamma_{L}\Gamma_{R}}{\Gamma}\int d\omega\left[  f_{R}\left(  \omega\right) -f_{L}\left(  \omega\right)\right]  \sum_{\sigma}\operatorname{Im}G_{\sigma}^{r}\left(\omega\right), \label{17}
\end{equation}
where $\Gamma=\sum_{\alpha}\Gamma_{\alpha}$ is the broadening of dot level $\varepsilon_{d\sigma}$ induced by the hybridization matrix elements $V_{\alpha}$ with $\Gamma_{\alpha}=\pi\rho\left\vert V_{\alpha}\right\vert^{2}$, and
\begin{equation}
G_{\sigma}^{r}\left(  \omega\right)  =-i\int dte^{i\omega t}\theta(t)\left\langle \left\{  d_{\sigma}\left(t\right), d_{\sigma}^{\dagger}(0)\right\}  \right\rangle \label{18}
\end{equation}
is the retarded Green's function (GF) of the dot electron. Here $\theta\left(t\right)$ is the Heaviside step function, and $f_{\alpha}\left(  \omega\right)=f(\omega-eV_\alpha)$ represents the Fermi distribution function in $\alpha$ lead and here $V_\alpha$ is the bias voltage of lead $\alpha$. In our practical calculation we adopt the symmetric bias. The differential conductance is defined as $G(V)=dI/dV$.

Due to the coupling between the photons and the electrons, the GF in Eq. (\ref{18}) is dressed by photon, which leads to sideband structure\cite{Mahan2000}. However, in experiment such sideband structure has not been observed in the bias voltage applied. Thus we here neglect such sideband effect but only consider the central band, which has a factor difference from the practical differential conductance, which does not affect the qualitative behavior of the differential conductance as a function of magnetic field.

\section{SLAVE-BOSON MEAN-FIELD THEORY}
To study the Kondo effect of the effective Hamiltonian obtained above, we express it in terms of the slave-boson operators\cite{Fang2013,Aguado2002}. In this way the fermionic operator of the dot is written as a combination of a pseudofermion and a boson operator: $d_\sigma=b^\dag f_\sigma$ where $f_\sigma$ is the pseudofermion which annihilates one ``occupied state" in the dot and $b^+$ is a boson operator which creates an ``empty state" in the dot. We consider the limit $\tilde U\rightarrow\infty$ so that the double occupancy of the dot is forbidden. Accordingly, a Lagrange multiplier $\gamma$ is introduced to enforce the constraint, $b^\dag b+\sum_\sigma f^\dag_\sigma f_\sigma=1$. Thus the Hamiltonian in the slave-boson language reads
\begin{eqnarray}\label{eqC1}
&&\tilde{H}_{SB}=\sum_{k\alpha\sigma}\varepsilon_{k\alpha\sigma}c_{k\alpha\sigma}^{\dag}c_{k\alpha\sigma}+\sum_{\sigma} \tilde\varepsilon_{d\sigma} f^\dag_\sigma f_\sigma \notag\\
&&+\sum_{k\alpha\sigma}\tilde{V}_{\alpha}\left[  b f_{\sigma}^{\dag}c_{k\alpha\sigma}+h.c.\right]+\gamma\left(  b^\dag b+\sum_\sigma f^\dag_\sigma f_\sigma-1\right).\label{eqC1}
\end{eqnarray}\label{eqB1}
We solve Eq. (\ref{eqC1}) within the mean-field approach, which is the leading order in a $1/N$ expansion. This approach sets the boson operator $b^\dag(b)$ to a classical, nonfluctuating value $r$, thereby neglecting charge fluctuations. The slave-boson mean field Hamiltonian is then given by
\begin{eqnarray}\label{eqB2}
&&\tilde{H}_{SBMF}=\sum_{k\alpha\sigma}\varepsilon_{k\alpha\sigma}c_{k\alpha\sigma}^{\dag}c_{k\alpha\sigma}+\sum_{\sigma}\left(  \tilde\varepsilon_{d\sigma}+\gamma\right) f^\dag_\sigma f_\sigma \notag\\
&&\hspace{1.2cm}\hspace{-0.7cm}+\sum_{k\alpha\sigma}r\tilde{V}_{\alpha}\left[  f_{\sigma}^{\dag}c_{k\alpha\sigma}+h.c.\right]+\gamma\left(  r^{2}-1\right), \label{eqB2}
\end{eqnarray}
where the two mean-field parameters $r$ and $\gamma$ have to be determined through their saddle-point equations which minimize the free energy $F =-k_BT \textrm{ln}[\textrm{Tr}(e^{\tilde H_{SBMF}/(k_BT )})]$, i.e.,
\begin{eqnarray}\label{eqC3}
\frac{\partial F}{\partial\gamma}=0,
\frac{\partial F}{\partial r} =0. \label{eqC3}
\end{eqnarray}\label{eqB2}
After some calculations, Eqs. (\ref{eqC3}) yields
\begin{eqnarray}\label{eqB3}
&& 1-r^{2}-\sum_{\sigma}\left\langle n_{f\sigma}\right\rangle  =0,\\
&& \sum_{k\alpha\sigma}r\tilde{V}_{\alpha}\left\langle f_{\sigma}^{\dag}c_{k\alpha\sigma}\right\rangle +\gamma r^{2} =0.
\end{eqnarray}\label{eqB3}
The thermal equilibrium averages above are related to the corresponding retarded GFs through the spectral theorem as follows
\begin{equation}
\langle BA\rangle=-\frac{1}{\pi}\int_{-\infty}^{\infty}d\omega f(\omega)\operatorname{Im}\langle\langle A,B\rangle\rangle,
\end{equation}
Using equation of motion method we obtain
\begin{eqnarray}
&&\langle\langle c_{k\alpha\sigma},f_{\sigma}^{\dag}\rangle\rangle =\frac{r\tilde{V}_{\alpha}}{w-\varepsilon_{k\alpha\sigma}}\left\langle
\left\langle f_{\sigma},f_{\sigma}^{\dag}\right\rangle \right\rangle , \label{eqC5}\\
&&\langle\langle f_{\sigma},f_{\sigma}^{\dag}\rangle\rangle  =\frac{1}{\omega-\tilde\varepsilon_{d\sigma}-\gamma-r^{2}\sum_{k\alpha}
\frac{\tilde{V}_{\alpha}^{2}}{\omega-\varepsilon_{k\alpha\sigma}}} \notag\\
&&\hspace{2cm} =\frac{1}{\omega-\tilde\varepsilon_{d\sigma}-\gamma+ir^{2}\tilde{\Gamma}}, \label{eqC6}
\end{eqnarray}
By using the spectral theorem we thus obtain
\begin{eqnarray}
&& \sum_{\sigma}\langle n_{f\sigma}\rangle  =-\frac{1}{\pi}\sum_{\sigma}\int d\omega f(\omega)  \operatorname{Im}\langle\langle f_{\sigma},f_{\sigma}^{\dag}\rangle\rangle,\label{eqC7}\\
&&\sum_{k\sigma\alpha}r\tilde{V}_{\alpha}\langle f_{\sigma}^{\dag}c_{k\sigma\alpha}\rangle  =\frac{r^{2}\tilde{\Gamma}}{\pi}\sum_{\sigma}\int d\omega f(\omega)  \operatorname{Re}\langle\langle f_{\sigma},f_{\sigma}^{\dag}\rangle\rangle.\label{eqC8}
\end{eqnarray}
Here we have made the approximation $-i\tilde{\Gamma}=\sum_{k\alpha}\frac{\tilde{V}_{\alpha}^{2}}{\omega-\varepsilon_{k\sigma\alpha}}$. For simplicity, we replace the symbol $\tilde \Gamma$ by $\Gamma$ in the main text. Substituting (\ref{eqC5}) and (\ref{eqC6}) into (\ref{eqC7}) and (\ref{eqC8}) we have
\begin{eqnarray}
&&1-r^{2}+\frac{1}{\pi}\sum_{\sigma}\int d\omega f(\omega)  \operatorname{Im}\langle\langle f_{\sigma},f_{\sigma}^{\dag}\rangle\rangle = 0, \label{eqB9}\\
&&\frac{\tilde{\Gamma}}{\pi}\sum_{\sigma}\int d\omega f(\omega)\operatorname{Re}\langle\langle f_{\sigma},f_{\sigma}^{\dag}\rangle\rangle +\gamma = 0. \label{eqB10}
\end{eqnarray}
After solving out $r$ and $\gamma$ we can calculate
\begin{equation}\label{eqB11}
\langle\langle d_{\sigma},d_{\sigma}^{\dag}\rangle\rangle=r^{2}\langle\langle f_{\sigma},f_{\sigma}^{\dag}\rangle\rangle =\frac{r^{2}}{\omega-\tilde\varepsilon_{d\sigma}-\gamma+ir^{2}\tilde\Gamma}.
\end{equation}
Now recalling Eq. (\ref{17}) we can turn to calculate the electronic current and then differential conductance.
\begin{figure}[h]
\begin{center}
\includegraphics[width=0.9\columnwidth]{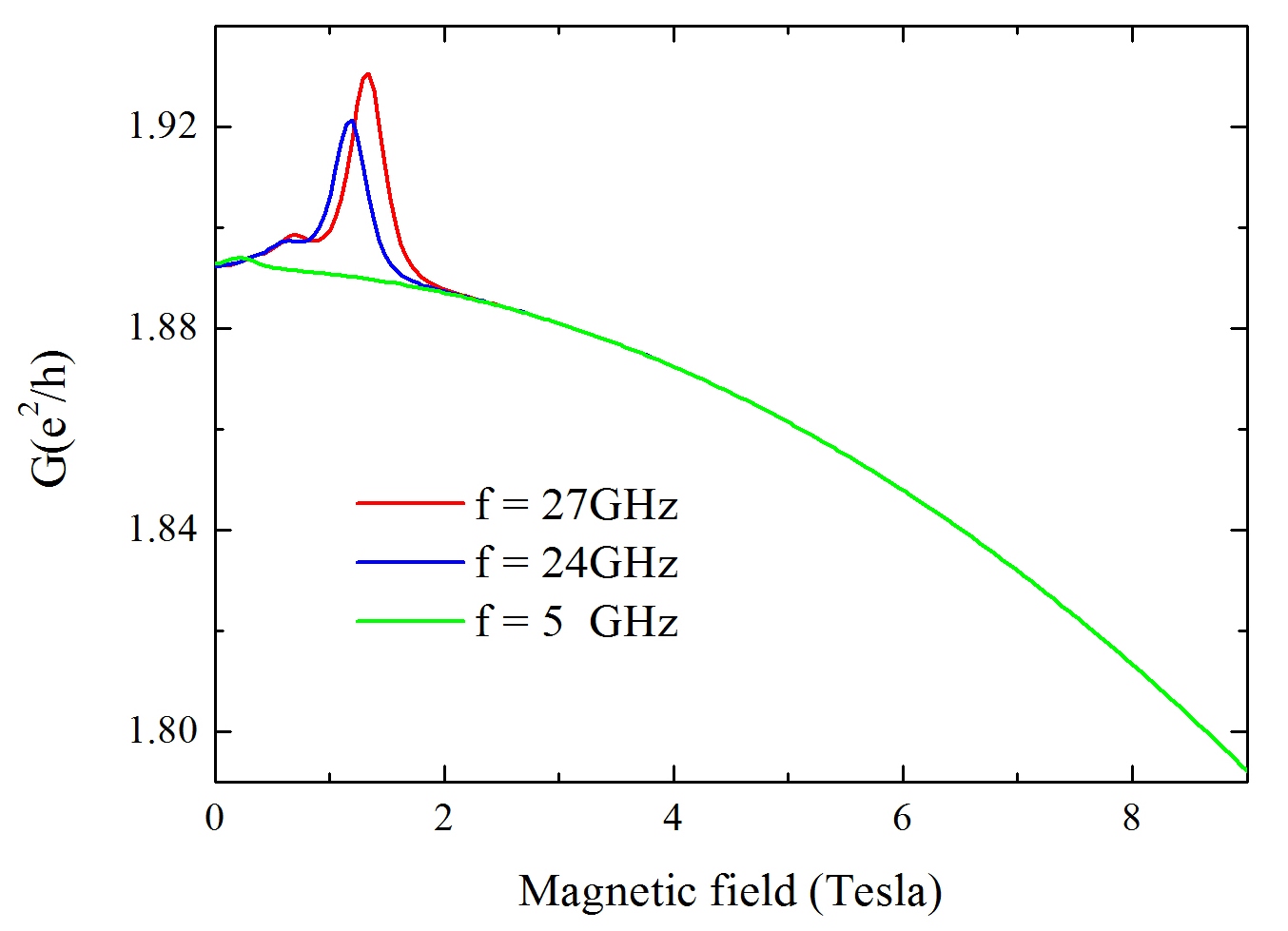}
\caption{(Color online) The differential conductance as a function of magnetic field under different microwave frequency. The resonance match is given by Eq. (\ref{eqd1}) and the parameters used are the same as those in Fig. 3(a) in the main text.}\label{fig5}
\end{center}
\end{figure}
\begin{table}[h]
\begin{tabular}{lcccccc}
\hline
\hline
$B(T)$ &0.98 &1.47 &1.68 &1.76 &1.86 &1.96\\
\hline
$G_0$ &0.3614	&0.3648	 &0.3558    &0.3573    &0.3600    &0.3712\\
\hline
$\rho_0$ &2.9450    &2.4264    &1.7424    &2.7515    &3.4200    &4.8905\\
\hline
$q$  &4.4958    &4.8379    &5.7301    &4.5688    &4.1269    &3.4676\\
\hline
$T_K(\mu eV)$ &157.856  &134.992  &112.161  &130.910  &149.935  &189.478\\
\hline
$\varepsilon_K(\mu eV)$ &-15.609  &-13.164   &-8.310  &-20.434  &-25.566  &-33.557\\
\hline
\hline
\end{tabular}\label{tab1}
\caption{Fitting parameters for different magnetic field $B$ applied.}
\end{table}
\section{Subpeak and/or SHOULDER STRUCTURES AT HALF AND ONE QUARTER}
As can be seen from the experimental results, there exist fine subpeak and/or shoulder structures of the differential conductance at the Fermi level as a function of magnetic field, which locate at about half and one quarter of the resonance magnetic field (see Fig.4 (d) of Ref.\cite{Hemingway2013}). In the paper \cite{Hemingway2013} this feature has not been pointed out. To reproduce these features in our match picture, one can consider phenomenologically the following resonance condition
\begin{equation}
\lambda = \sum_{i=0}^{2}\lambda_{i} \delta(\omega_p - \Omega_{\lambda,i}\Delta\varepsilon),\label{eqd1}
\end{equation}
where $\lambda_i = 2^{-i}\lambda_0$ is assumed and $\lambda_0$ is considered to be dependent of the microwave frequency. In addition, the renormalization factor $\Omega_{\lambda,i} = 2^i\Omega_\lambda$ is also assumed. Here we do not know essential physics under which these subpeaks and/or shoulder occur. Anyway, these features can be captured phenomenologically by our match picture, as shown in Fig. \ref{fig5}.

\section{FITTING PARAMETERS OF FANO FORMULA}
In Table \ref{tab1} we present the fitting parameters introduced in Eqs. (7) and (8) in the main text for the experimental data given in Fig. 4(a) and (b) for different magnetic fields. The fitting result is shown in Fig. 4(a) in the main text. The Kondo temperature obtained are in reasonable agreement with the experiment data, namely, about 1K.